\documentclass[10pt,sigconf,natbib=false]{acmart} 

\usepackage{epsfig}
\usepackage[tight,footnotesize]{subfigure} 
\usepackage{caption} 
\usepackage{epstopdf}
\usepackage{balance}
\usepackage{minipage}

\usepackage{booktabs} 

\setcopyright{none}

\acmConference[]{}
\acmPrice{ }

\renewcommand\footnotetextcopyrightpermission[1]{} 
\usepackage{color}
\usepackage{xcolor}
\usepackage{colortbl}

\definecolor{Gray}{gray}{0.9}

\definecolor{linkcol}{rgb}{0,0,0.75}
\definecolor{citecol}{rgb}{0,0,0}
\definecolor{urlcol}{rgb}{0,0,0}

\usepackage{hyperref}

\usepackage{breakurl}

\usepackage{booktabs}
\usepackage{tabularx}

\usepackage{array} 
\newcolumntype{H}{>{\setbox0=\hbox\bgroup}c<{\egroup}@{}}

\usepackage{comment}
\long\def\comment#1{}

\newcommand{\remove}[1]{}

\includecomment{emergencydroppable} 
\excludecomment{droppingdetailforspace} 

\providecommand{\ie}{\emph{i.e.,} }
\providecommand{\eg}{\emph{e.g.,} }

\usepackage{amssymb}
\usepackage{amsmath}
\usepackage{mathrsfs}
\usepackage{mathbbol}
\usepackage{bm}			

\usepackage{amsthm} 

\newcommand{\myitem}[1]{\vspace*{0.04in}\noindent\textbf{#1}}

\usepackage[normalem]{ulem}

\newcommand{\cm}[1]{\mbox{\hspace{1cm}\text{#1}\hspace{1cm}}}

\usepackage{cleveref}
\crefformat{footnote}{#2\footnotemark[#1]#3}

\usepackage{verbatim}

\usepackage{listings}

\usepackage{enumitem}
\setlist[description]{leftmargin=*}

\usepackage{cite}

\usepackage{xspace}
\newcommand{\artemis}[0]{\textsf{ARTEMIS}\xspace} 

\def\papertitle{
A Survey among Network Operators \\on BGP Prefix Hijacking
}

\hypersetup{
	pdfauthor = {},
	pdftitle = {\papertitle},
	pdfcreator = {LaTeX with hyperref package},
	pdfproducer = {dvips + ps2pdf}
}

\begin{document}

\title{\papertitle}

\newcommand{\inst}[1]{$^{#1}$}
\author{
\hspace{-10pt}
Pavlos~Sermpezis\inst{1},
Vasileios~Kotronis\inst{1}, 
Alberto~Dainotti\inst{2},
Xenofontas~Dimitropoulos\inst{1,3}
\\[4pt]
$^1$FORTH-ICS
\hspace{10pt}
$^2$CAIDA, UC San Diego
\hspace{10pt}
$^3$University of Crete
}

\renewcommand{\shortauthors}{P. Sermpezis et al.}



\begin{abstract}
BGP prefix hijacking is a threat to Internet operators and users.
Several mechanisms or modifications to BGP that protect the Internet
against it have been proposed. However, the reality is that most
operators have not deployed them and are reluctant to do so in the
near future. Instead, they rely on basic - and often inefficient - proactive defenses to reduce the impact of hijacking events, or on detection based on third party services and reactive approaches that might take up to several hours. 
In this work, we present the results of a survey we conducted among 75 network operators to study: (a) the operators' awareness of BGP prefix hijacking attacks, (b) presently used defenses (if any) against BGP prefix hijacking, (c) the willingness to adopt new defense mechanisms, and (d) reasons that may hinder the deployment of BGP prefix hijacking defenses. We expect the findings of this survey to increase the understanding of existing BGP hijacking defenses and the needs of network operators, as well as contribute towards designing new defense mechanisms that satisfy the requirements of the operators.

\end{abstract}

\maketitle

\section{Introduction}
\textbf{BGP prefix hijacking 101.} 
Autonomous Systems (ASes) use the Border Gateway Protocol (BGP)~\cite{BGPv4} 
to advertise address space (as IPv4/IPv6 network prefixes) and establish inter-domain routes in the Internet. 
BGP is a distributed protocol, lacking authentication of advertised routes. As a
result, an AS is able to advertise illegitimate routes for IP prefixes it does 
not own. These advertisements propagate and ``pollute'' many ASes, or even the 
entire Internet, affecting service availability, integrity, and confidentiality of
communications.
This phenomenon, called \textit{BGP prefix hijacking}, is frequently
observed~\cite{Vervier-mind-blocks-NDSS-2015}, and can be caused by
router misconfigurations~\cite{hijack-YouTube,hijack-ChinaTelecom} or
malicious attacks~\cite{hijack-BitCoins,Ramachandran-BGP-spammers-CCR-2006,
Vervier-mind-blocks-NDSS-2015}. 

\myitem{Current defenses are not sufficient.} 
Currently, networks rely on \textit{practical reactive mechanisms} to
defend against prefix hijacking, since \textit{proactive mechanisms} such as RPKI~\cite{Kent-secure-BGP-JSAC-2000,Subramanian-listen-whisper-NSDI-2004,bgpsec-specification-2015,rpki-rfc,Karlin-PGBGP-ICNP-2006}
are fully efficient only when globally deployed, and operators are
reluctant to deploy them due to associated technical and financial
costs~\cite{lychev-2013-sigcomm,cooper-2013-hotnets,george-2014-adventures,Goldberg-why-so-long-ACM-Comm-2014,matsumoto-2017-privsec}.
Reactive mechanisms mainly operate in two stages: \textit{detection} (\eg based on monitoring data) and \textit{mitigation} (\eg based on local network actions, such as originating BGP advertisements) of the hijack.
The speed of the reactive defenses is crucial; even short-lived events can have severe consequences~\cite{hijack-BitCoins}. However, the reality shows that, currently, hijacking events are not quickly mitigated. 
For instance, back in 2008, a hijacking event affected YouTube's
prefixes and disrupted its services for $2$ hours~\cite{ripe-pakistan}. 
More recently, in Sep.~2016, BackConnect (AS203959) hijacked, at different times, several ASes; 
the events lasted for several hours~\cite{backconnect-hijack-2016}. 
In Jan.~2017, the Iranian state telecom TIC hijacked disparate pornographic 
websites for more than a day~\cite{iranian-hijack-2017}. 
In Apr.~2017, financial services, like Visa and Mastercard, and security companies, 
like Symantec, were hijacked by a Russian company for seven minutes~\cite{russian-hijack-2017}. 

\myitem{Survey motivation and contributions.}
To surpass existing shortcomings and achieve a swift and efficient resolution of hijacking events, new defense approaches that fit the needs and requirements of the operators are needed. To this end, we launched a survey~\cite{google-survey-hijacking} to increase the understanding of currently used BGP hijacking defenses, and to receive feedback directly from network operators about their needs. The main motivation for this survey was to propose and design such a defense approach that matches
the community needs; in fact, we acquired valuable information from this survey that helped us design a defense system called \artemis~\cite{artemis-techrep-2017}. 

However, the findings of the survey are more general and can be beneficial for both researchers and operators. Researchers can evaluate the severity of the problem of BGP prefix hijacking as it is seen from the operator community, and investigate new defense mechanisms capitalizing on current operational practices. Operators can be informed about the trends in the BGP prefix hijacking issue and the employed defenses, provide valuable feedback to the network community themselves, and adjust accordingly the way they manage and protect their networks against hijacks.

\myitem{Structure.}
In Section~\ref{sec:survey-questions} we present the questions of the survey, and in Section~\ref{sec:survey-results} we discuss the main findings and their implications. The detailed results are presented in Figures~\ref{fig:questions-section-1},~\ref{fig:questions-section-2}, and~\ref{fig:questions-section-3}.

\section{Survey Profile and Questions}\label{sec:survey-questions}
\begin{figure*}[h]
\centering
\begin{minipage}[b]{0.23\linewidth}
\subfigure[Q1: Which term(s) would best characterize your organization?]{\includegraphics[width = 1\linewidth]{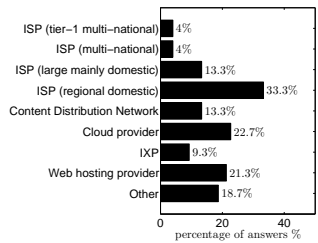}\label{fig:survey-q1}}
\end{minipage}
\hspace{0.23\linewidth}
\begin{minipage}[b]{0.23\linewidth}
\subfigure[Q2: In which continent(s) does your company operate?]{\includegraphics[width = 1\linewidth]{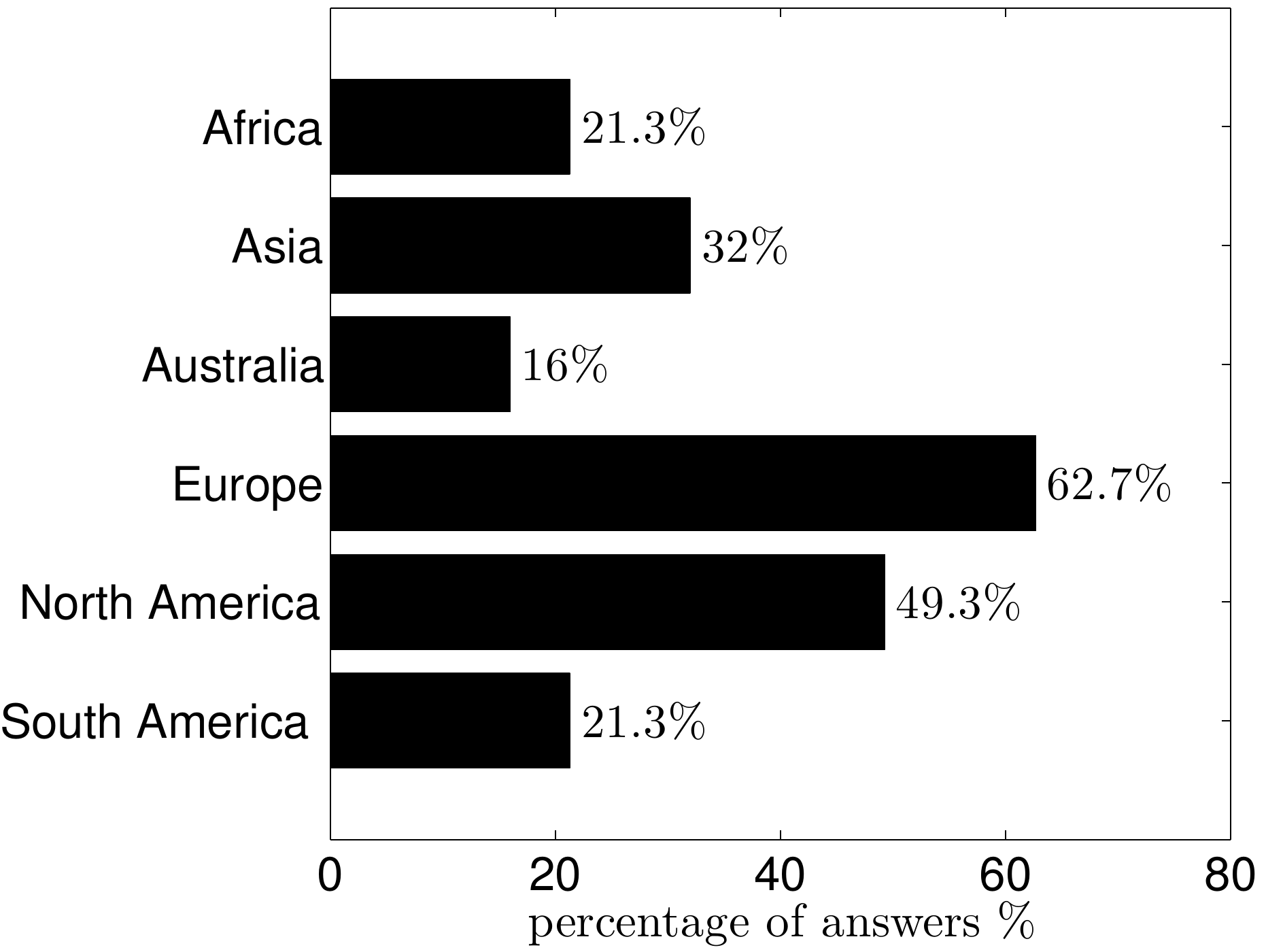}\label{fig:survey-q2}}
\end{minipage}

\vspace{0.5\baselineskip}

\begin{minipage}[b]{0.46\linewidth}
\subfigure[Q3: In which country(-ies) does your company operate? (optional)]{\includegraphics[height = 0.38\linewidth]{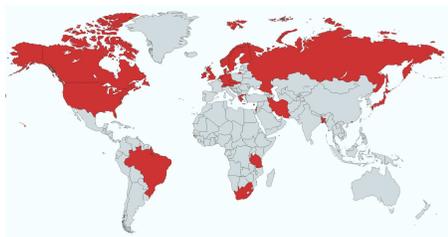}\label{fig:survey-q3}}
\end{minipage}
\begin{minipage}[b]{0.23\linewidth}
\subfigure[Q4: What is your position in your company?]{\includegraphics[width = 1\linewidth]{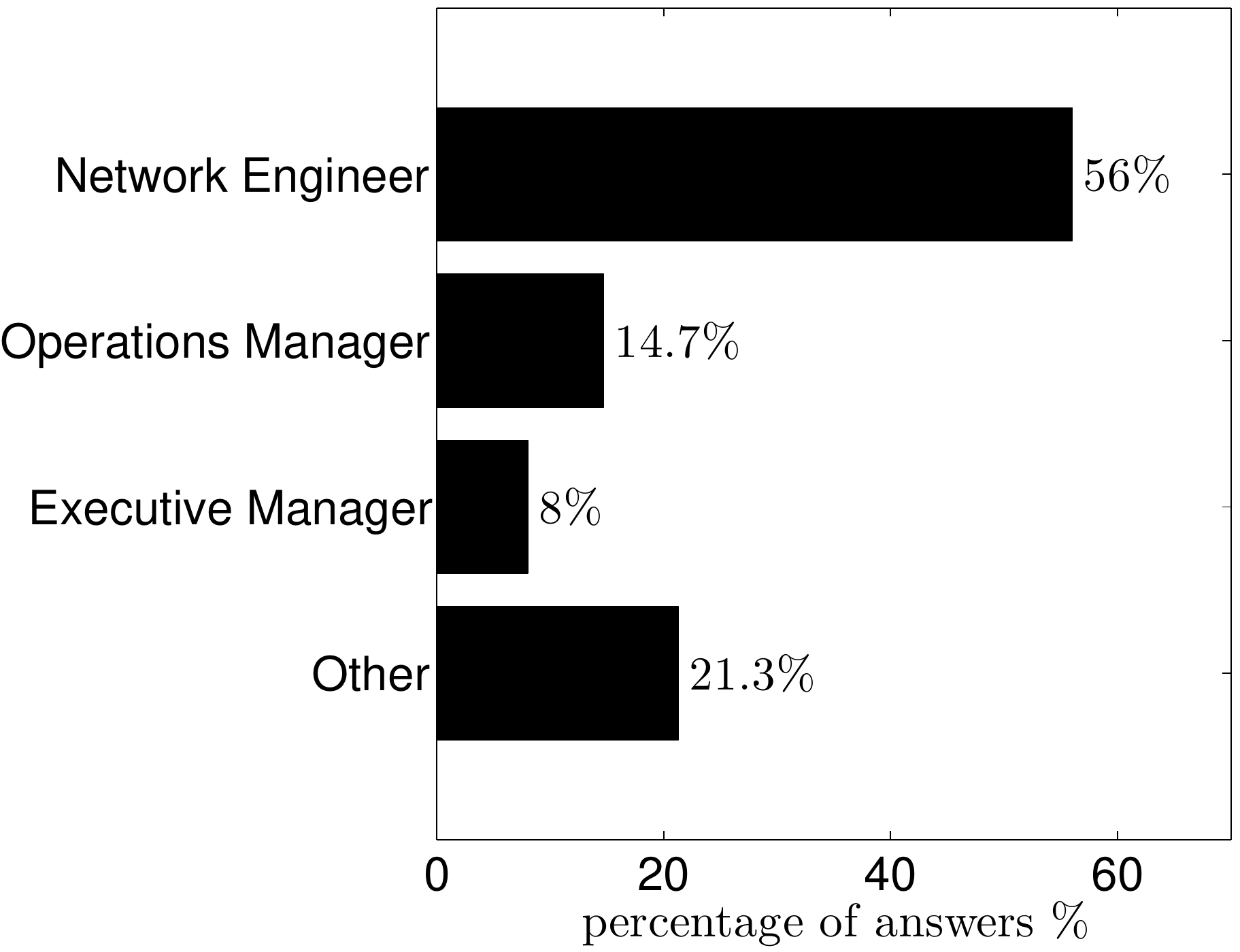}\label{fig:survey-q4}}
\end{minipage}
\caption{Survey results -- Information about the participants and their organizations}
\label{fig:questions-section-1}

\vspace{1\baselineskip}

\end{figure*}

We launched a
survey~\cite{google-survey-hijacking} on network operators' mailing
lists, such as NANOG and RIPE. The survey is anonymous and comprises
21 questions studying \textit{(a)} the operators' awareness of BGP
prefix hijacking attacks, \textit{(b)} presently used defenses against
BGP prefix hijacking, \textit{(c)} the willingness to adopt new defense
mechanisms, and \textit{(d)} reasons that may hinder the deployment of BGP
prefix hijacking defenses. We received answers from 75 participants
operating a broad variety of networks (Fig.~\ref{fig:survey-q1}) all
over the world (Figs.~\ref{fig:survey-q2} and~\ref{fig:survey-q3}),
working at different positions (Fig.~\ref{fig:survey-q4}).

The survey/questionnaire is composed of three parts.

\myitem{(1) Information about the participants and their organizations (4 questions)}. In the first part we ask the participants to provide information about the type (\eg ISP, CDN, IXP) and location of their organization, as well as their work position in the organization. The questions and results are presented in Fig.~\ref{fig:questions-section-1}.

\myitem{(2) Knowledge and Experience with BGP Prefix Hijacking (6 questions).} The second part consists of questions related to the participants' awareness and concern about BGP prefix hijacking, including their experience with past hijacking events on their networks. The questions and results are presented in Fig.~\ref{fig:questions-section-2}.

\myitem{(3) Defenses against BGP Prefix Hijacking (11 questions).} The last part asks the participants about (i) the defenses they use (if any) against BGP prefix hijacking, such as RPKI, (ii) how they detect and mitigate a hijacking event affecting their prefixes, and (iii) the characteristics they consider desirable (or not) in a future defense (detection/mitigation) system. The questions and results are presented in Fig.~\ref{fig:questions-section-3}.

\section{Survey Results}\label{sec:survey-results}
We classify the survey findings in 4 categories, which we present in the following sections:
(i) evaluation of impact of hijacks (Section~\ref{sec:survey-impact}),
(ii) general information about current defense mechanisms employed against hijacks (Section~\ref{sec:survey-defences}),
(iii) specific information on the detection and mitigation stages in today's operations (Section~\ref{sec:survey-detect-mitigate}), and
(iv) requirements posed on new mitigation mechanisms (\eg involving outsourcing defense functionality to third parties), as well as the willingness of operators to adopt them (Section~\ref{sec:survey-new-mitigation}).

\subsection{Impact of Hijacks}\label{sec:survey-impact}

\begin{figure*}[h]
\centering
\begin{minipage}[b]{0.23\linewidth}
\subfigure[Q5: Do you know what BGP prefix hijacking is and how it can happen?]{\includegraphics[width = 1\linewidth]{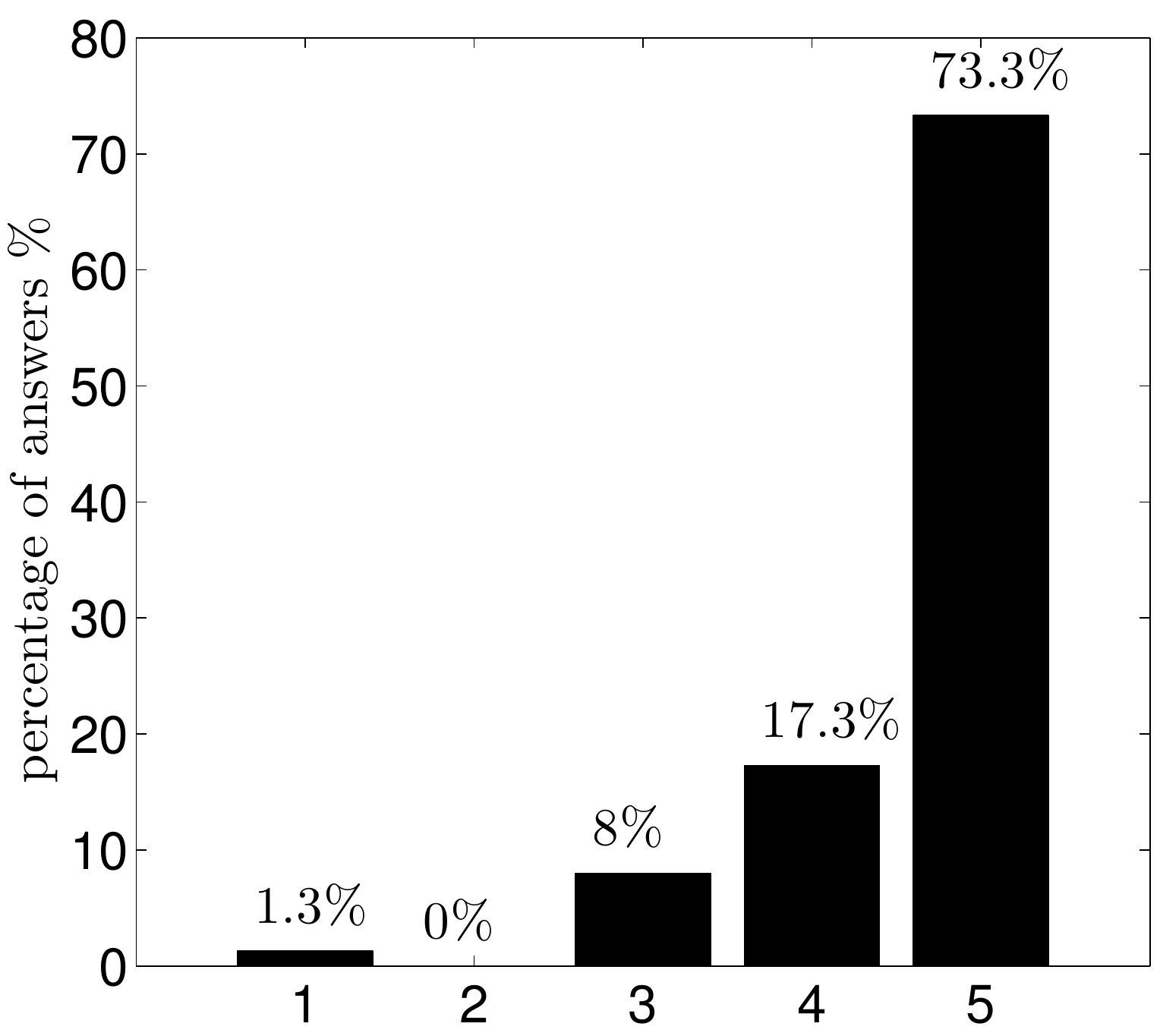}\label{fig:survey-q5}}
\end{minipage}
\hspace{0.1\linewidth}
\begin{minipage}[b]{0.23\linewidth}
\subfigure[Q6: How concerned are you about BGP prefix hijacking incidents on the Internet?]{\includegraphics[width = 1\linewidth]{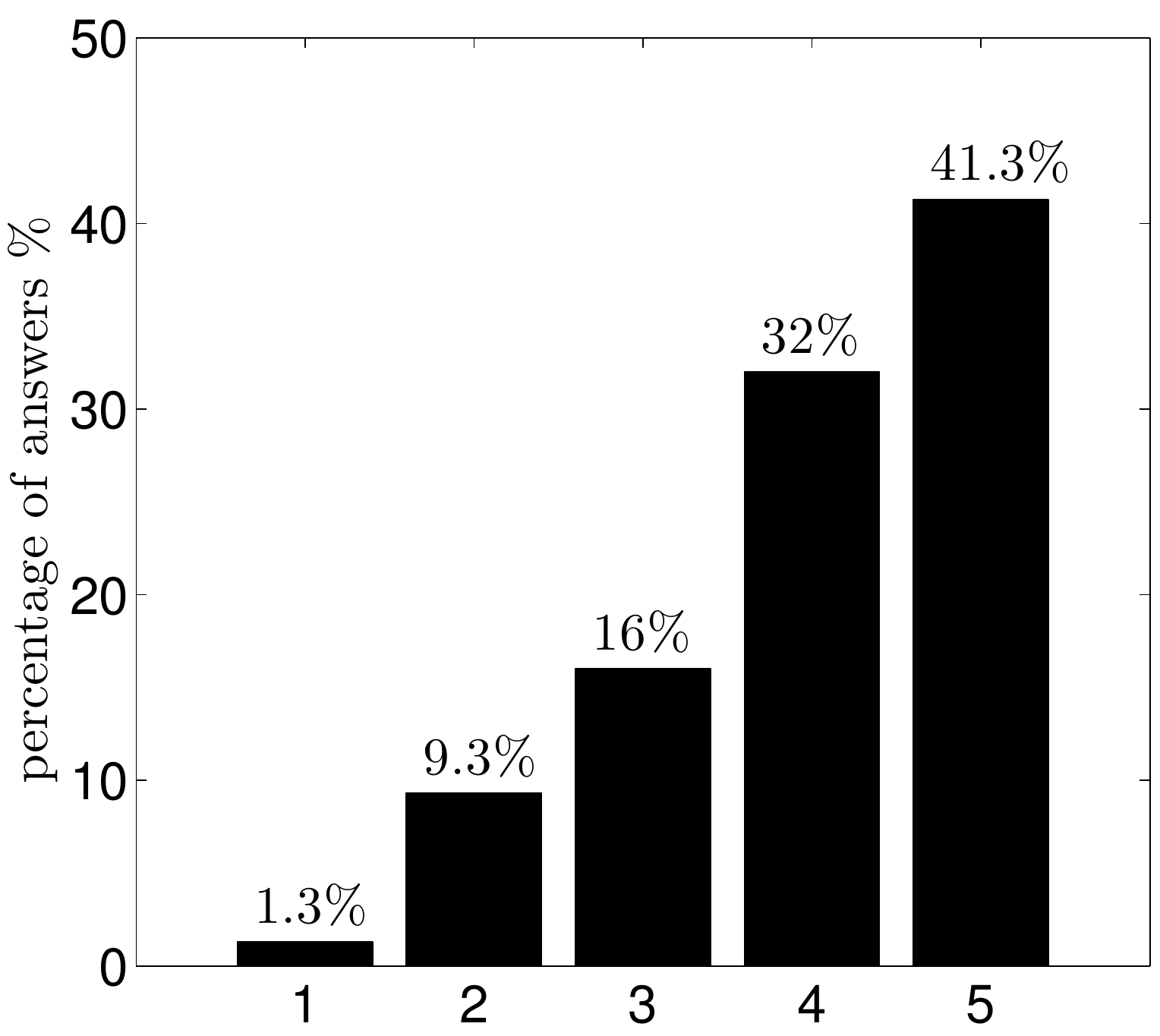}\label{fig:survey-q6}}
\end{minipage}
\hspace{0.1\linewidth}
\begin{minipage}[b]{0.23\linewidth}
\subfigure[Q7: Are you concerned that your network may be a victim of a BGP prefix hijacking incident in the future?]{\includegraphics[width = 1\linewidth]{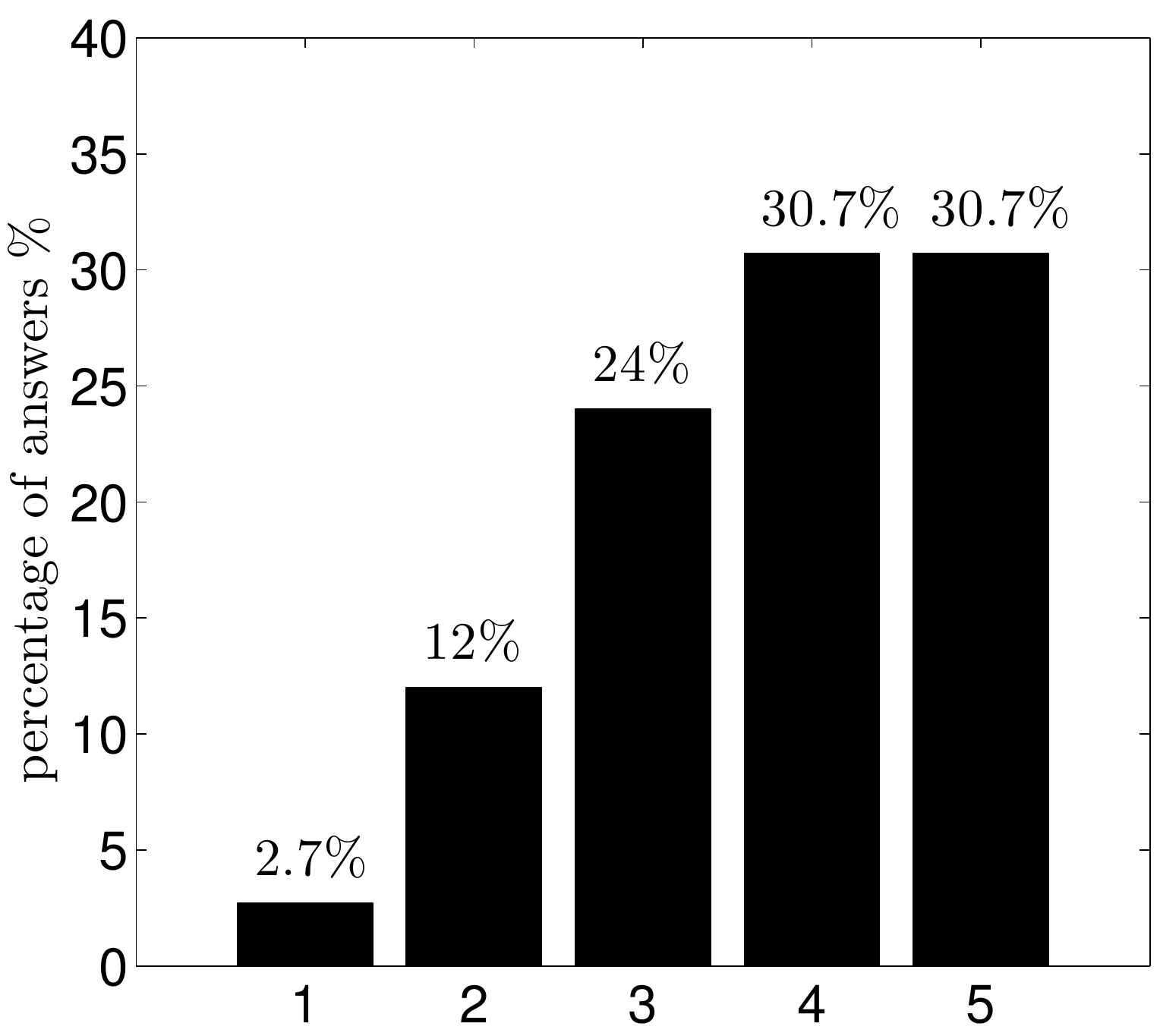}\label{fig:survey-q7}}
\end{minipage}

\vspace{0.5\baselineskip}

\begin{minipage}{0.23\linewidth}
\vspace{-1cm}
\subfigure[Q8: How severe do you consider the potential impact of a BGP prefix hijacking against your network?]{
\centering
\begin{footnotesize}
\vspace{-1cm}
\begin{tabular}{|c|ccc|}
\hline
{}& {no}& {}& {}\\
{}& {\hspace{-0.1cm}impact}& {\hspace{-0.3cm}$\sim$min.}& {\hspace{-0.3cm}$\sim$hours}\\
\hline
{few }&{}&{}&{}\\
{services/}& {0\%}& {~9.3\%}& {28.0\%}\\
{clients}&{}&{}&{}\\
\hline
{many}&{}&{}&{}\\
{services/}& {0\%}& {~9.3\%}& {48.0\%}\\
{clients}&{}&{}&{}\\
\hline
\end{tabular}
\end{footnotesize}
\label{table:survey-q8}
}
\end{minipage}
\hspace{0.1\linewidth}
\begin{minipage}[b]{0.23\linewidth}
\subfigure[Q9: Has your organization been a victim of a BGP prefix hijacking incident in the past?]{\includegraphics[width = 1\linewidth]{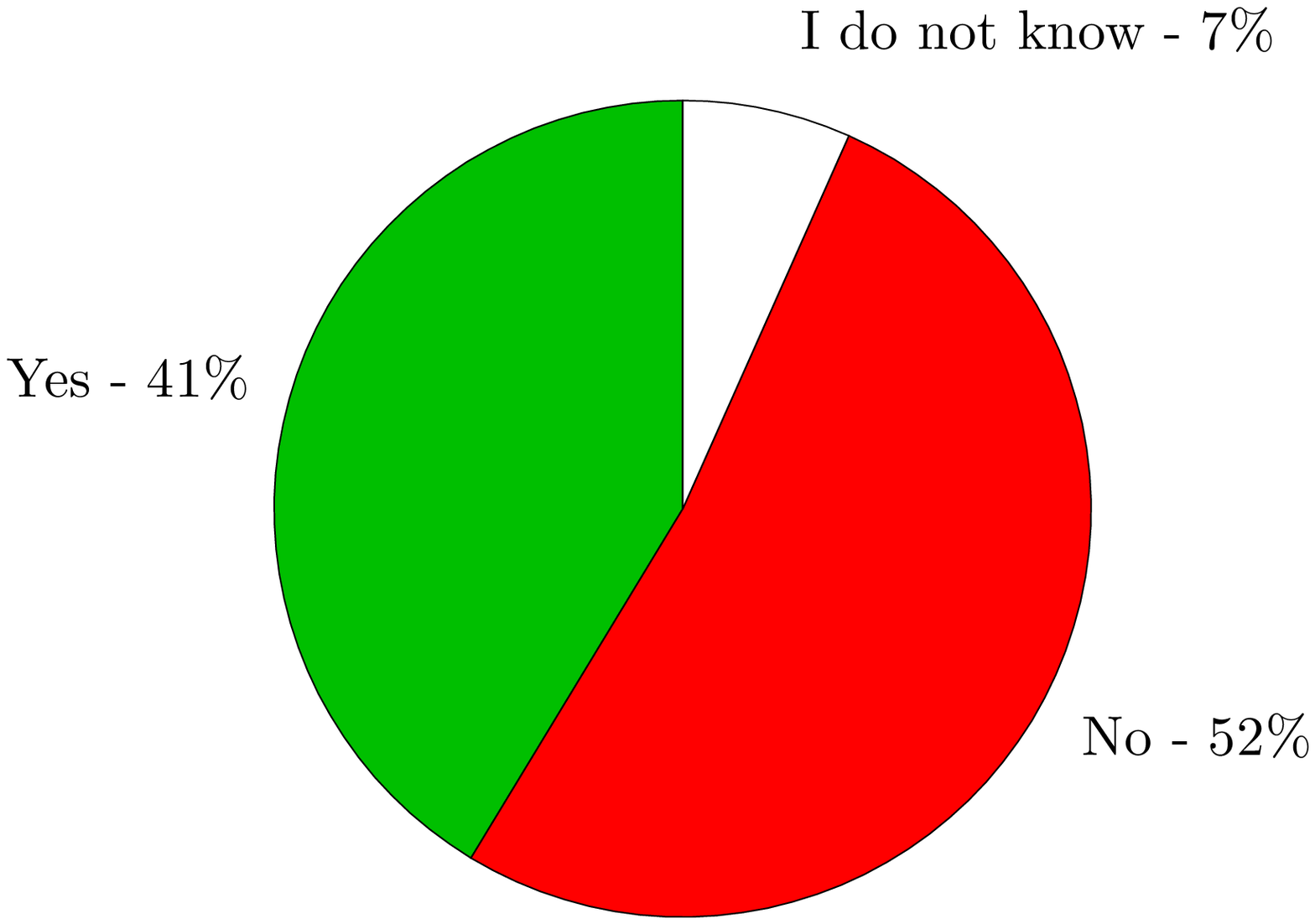}\label{fig:survey-q9}}
\end{minipage}
\hspace{0.1\linewidth}
\begin{minipage}[b]{0.23\linewidth}
\subfigure[Q10: If your organization was a victim of a BGP prefix hijacking incident, for how long was your network affected? (optional)]{\includegraphics[width = 1\linewidth]{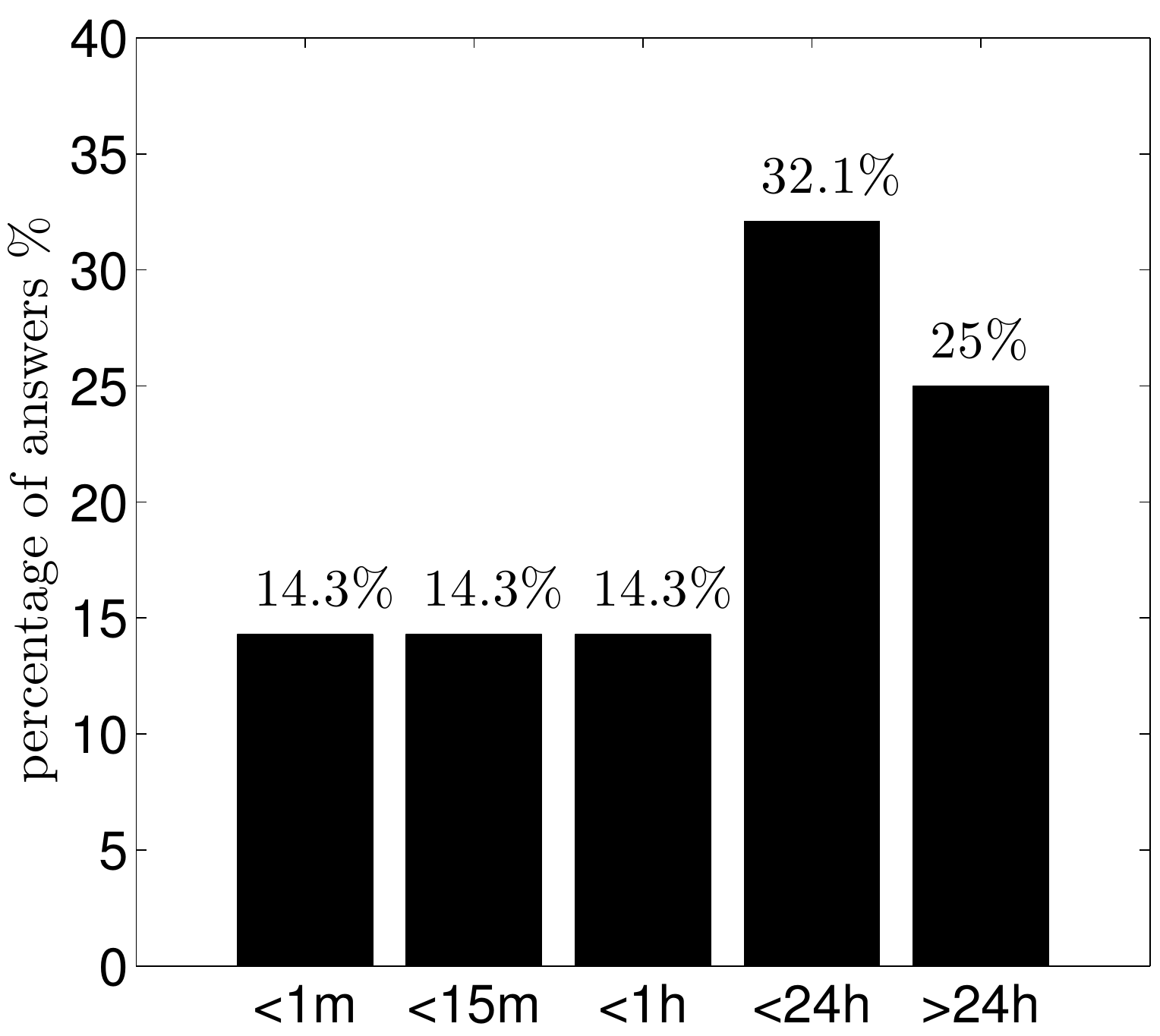}\label{fig:survey-q10}}
\end{minipage}
\caption{Survey results -- Knowledge and Experience with BGP Prefix Hijacking}
\label{fig:questions-section-2}

\end{figure*}

\myitem{BGP prefix hijacking is a real threat and concerns network operators.} More than 40\% of the operators reported that their organization has been a victim of a hijack in the past (Fig.~\ref{fig:survey-q9}). However, the vast majority is concerned about BGP prefix hijacking in the Internet (Fig.~\ref{fig:survey-q6}) and its potential impact on their own networks (Fig.~\ref{fig:survey-q7}). Almost all operators are knowledgeable on the issue of hijacks and the involved mechanisms (Fig.~\ref{fig:survey-q6}).

\myitem{Hijacks have a severe and lasting impact.} Operators evaluate the impact of a potential hijack targeting their network (in terms of \textit{duration} and \textit{number of disrupted services}) as shown in Fig.~\ref{table:survey-q8}. 
The vast majority (76\%) expects the impact of a hijack to last for a long time (few hours or more), while opinions are divided on whether the hijack will affect a few or many of their services/clients, indicating that there are concerns
both for extended (\eg route leaks) and limited/targeted (\eg malicious attacks) hijacks. Moreover, their past experience (Fig.~\ref{fig:survey-q10}) shows that most hijacks indeed lasted long: more than 57\% of hijacks lasted more than an hour, while 25\% lasted \textit{more than a day}; around 28\% are short-term hijacks, lasting a few minutes (14.3\%) or seconds (14.3\%).

\subsection{Defenses against Hijacks}\label{sec:survey-defences}

\myitem{RPKI deployment is limited.} In accordance with previous studies~\cite{RPKI-deployment-2016}, most of the network operators (71\%) answered that they have not deployed RPKI as a proactive defense mechanism in their networks (Fig.~\ref{fig:survey-q11}); very few (12\%) use the full functionality of RPKI (Route Origin Authorisation - ROA and Route Origin Validation - ROV). There are various reasons for this, as shown in Fig.~\ref{fig:survey-q12}; deployment lags mainly due to RPKI's \textit{limited adoption} and \textit{little security benefits}, but also due to the increased \textit{CAPEX and OPEX costs}, and increased \textit{complexity} and \textit{processing overhead} associated with the protocol mechanisms. Therefore, about 60\% of the operators (Fig.~\ref{fig:survey-q13}) resort to other mechanisms and practical defenses to protect their networks against BGP hijacks. 

\myitem{Practical defenses include route filtering, extensive peering, and de-aggregation.} The responses to the (optional) question ``what other defense mechanisms are used by networks'' are shown in Fig.~\ref{fig:survey-q14}. The majority of the participants, \ie 17 networks (among those who provided answers for this optional question), use \textit{route filtering} as a proactive defense to protect their own and their customers' prefixes from being hijacked. \textit{Route filtering} is implemented in various ways (based on their answers) including for example: prefix origin (e.g., from IRR records) or AS-path filtering; filtering at edge routers (with customers/peers) or route servers (at IXPs). Less popular approaches are \textit{anycast} 
(2 answers) and \textit{prefix de-aggregation} (4 answers). Finally, 5 operators (from CDNs or tier-1 networks) mention that they peer with many other networks extensively; this helps them protect their networks from hijacking events (\ie by reducing their impact).

\subsection{Detection and Mitigation of Hijacks.}\label{sec:survey-detect-mitigate}

\myitem{Hijack detection mainly relies on third parties.} The majority of networks (61.3\%) use a third party detection service, which notifies them about hijacking incidents against their prefixes (Fig.~\ref{fig:survey-q15}). BGPmon~\cite{commercial-bgpmon} is the most popular detection service, according to Fig~\ref{fig:survey-q16}. The satisfaction of operators from third parties generally varies a lot; some use them because they are satisfied and others because there are no alternatives (\eg it is not possible to develop their own detection service)\footnote{This variation can be observed in the following examples from the detailed answers in our survey:\\
\noindent\textit{``pretty happy with it'', ``no issues so far'', ``it works fine [..], but is relatively limited'', ``I hate it'', ``It's ok'', `` it seems to work quite well the few times I have needed it'', ``Better than nothing, but a lot of false alerts'', ``It rules!'', ``It is very noisy because it does not know a damn thing about IXP route servers'', ``Not great''.}
}. 
Moreover, 17.3\% of networks also practically rely on third parties, since they expect to get notified about a hijack by receiving notification from colleagues, clients, mailing lists, etc. In total, 78.6\% of the networks rely on third parties for the detection of hijacks against their prefixes. About one third of the networks have deployed a local hijack detection mechanism (\eg by monitoring the disruption of their services)\footnote{Among the ``other'' answers, a high percentage of answers relates to the observation of -or, reception of complaints about- disruption in their services.}. Finally, a non-negligible percentage of 8\% would probably not learn about a hijack. 

\myitem{Mitigating through de-aggregation and contacting other networks.} Asking operators what would be the countermeasures they would take to mitigate a prefix hijacking\footnote{Note that for this question, we provided the choices, based on the answers received in a preliminary version of our survey; operators could have answered in a different way if this was a completely open question.}, the majority (62.7\%; Fig.~\ref{fig:survey-q17}) responded that they would announce more specific prefixes (de-aggregation) and contact the offending network (i.e., the hijacker) or its providers. 5.3\% would follow \textit{only} the former approach (de-aggregation) and 25.3\% \textit{only} the latter (contacting other operators). This indicates that although de-aggregation is not widely used currently (see Fig.~\ref{fig:survey-q14}), operators still find it a good solution and are willing to proceed to similar actions after a hijacking event--affecting their--networks has taken place.

\subsection{New Mitigation Mechanisms}\label{sec:survey-new-mitigation}

\begin{figure*}[h]
\centering
\begin{minipage}[b]{0.23\linewidth}
\subfigure[Q11: Do you use RPKI in your network?]{\includegraphics[width = 1\linewidth]{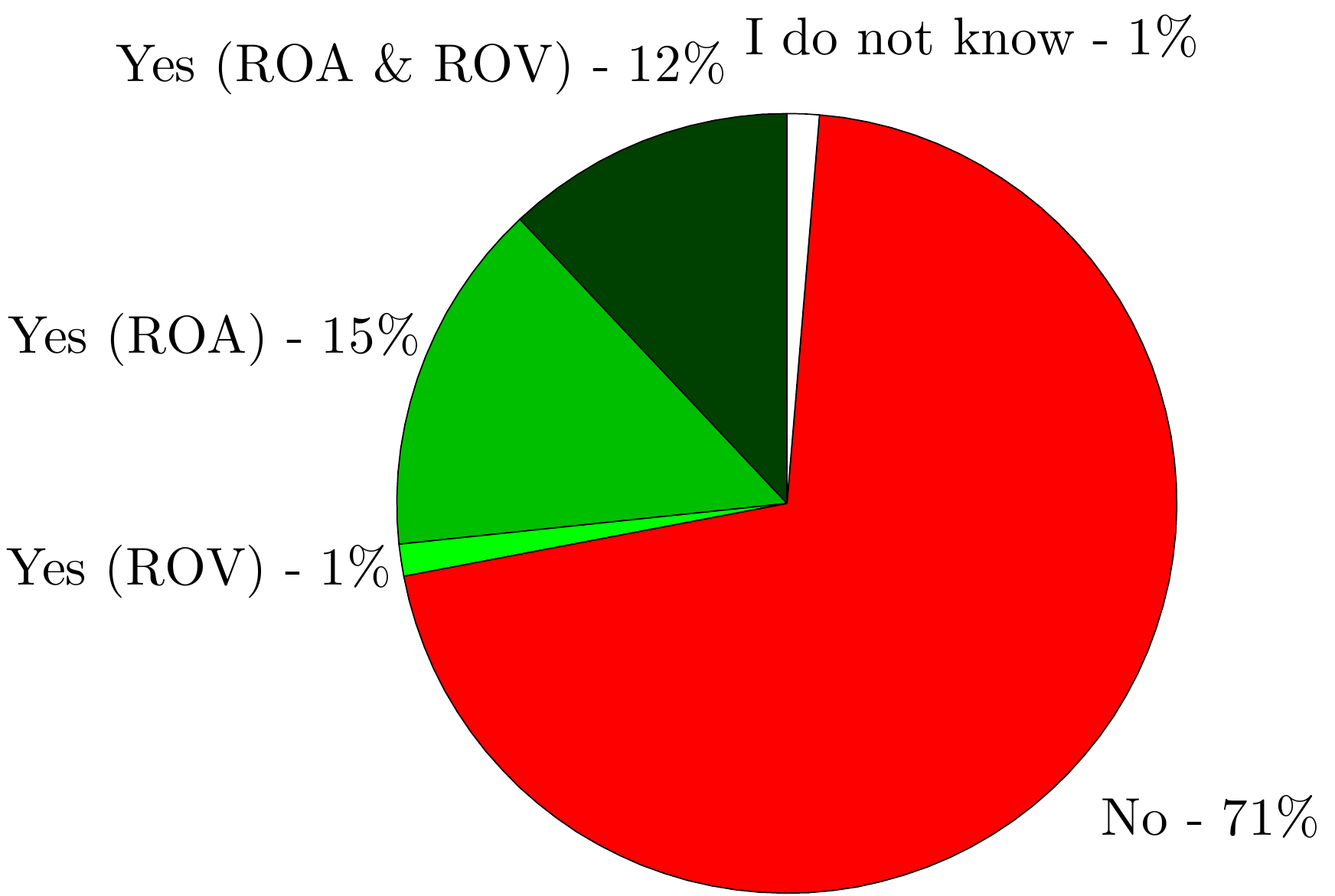}\label{fig:survey-q11}}
\end{minipage}
\hspace{0.01\linewidth}
\begin{minipage}[b]{0.23\linewidth}
\subfigure[Q12: If no (in Q11), what are the main reasons for not using RPKI? (optional)]{\includegraphics[width = 1\linewidth]{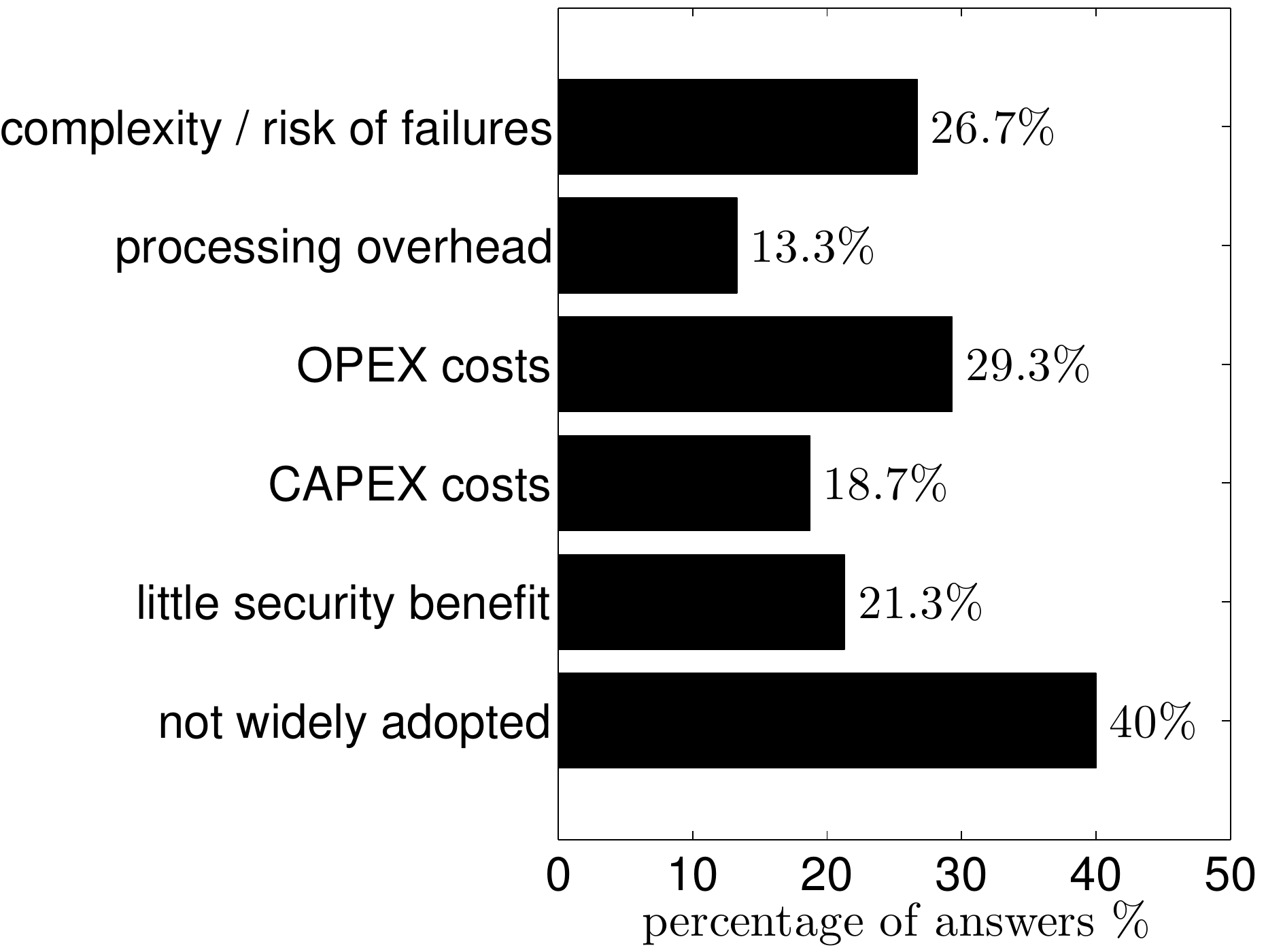}\label{fig:survey-q12}}
\end{minipage}
\hspace{0.01\linewidth}
\begin{minipage}[b]{0.23\linewidth}
\subfigure[Q13: Do you use in your network any other defense mechanisms (other than RPKI) that protect your/others' prefixes from BGP prefix hijacking?]{\includegraphics[width = 1\linewidth]{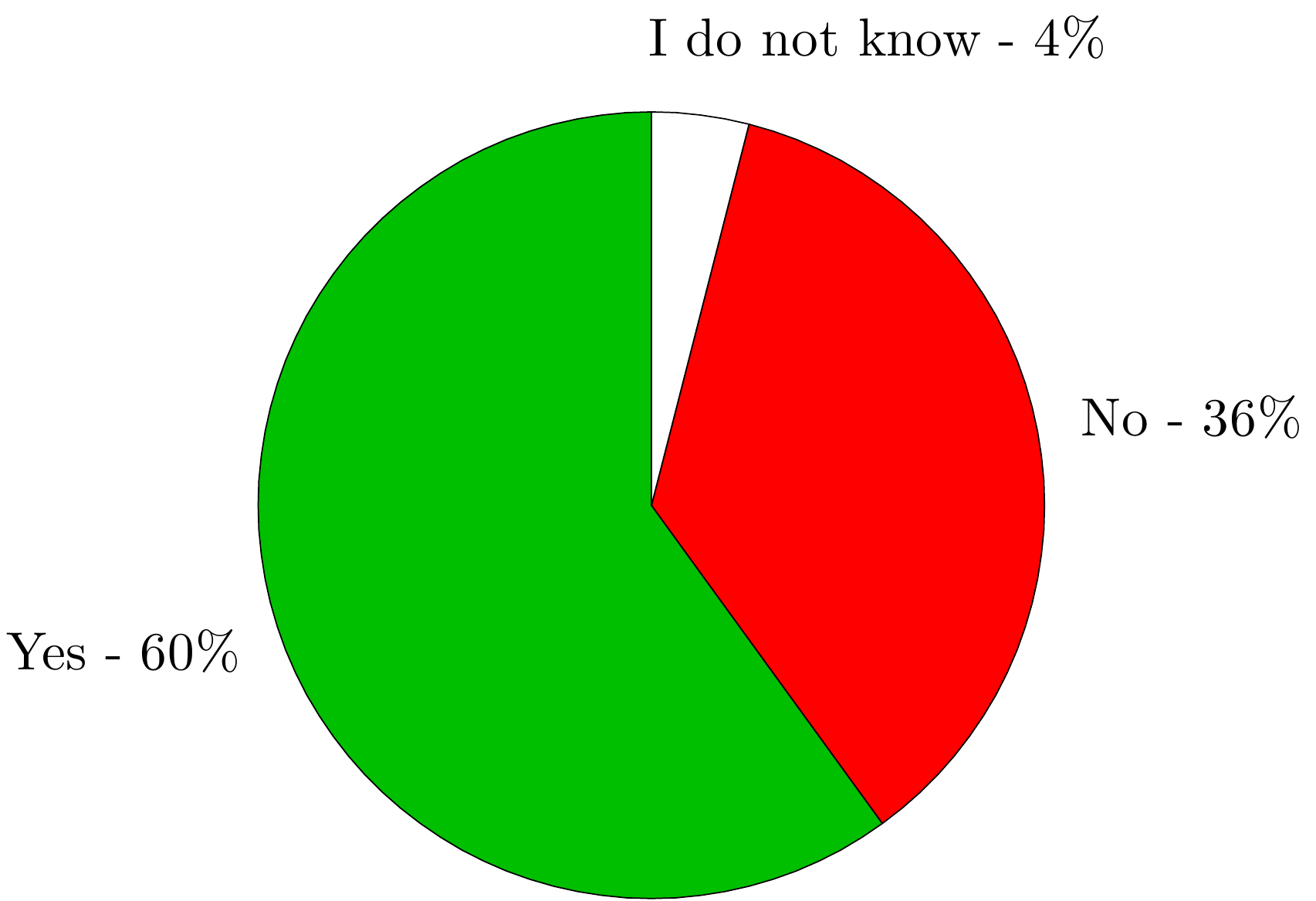}\label{fig:survey-q13}}
\end{minipage}
\hspace{0.01\linewidth}
\begin{minipage}[b]{0.23\linewidth}
\subfigure[Q14: If yes (in Q13), what mechanisms do you use? Could you provide a brief description? (open question/answers - optional)]{\includegraphics[width = 1\linewidth]{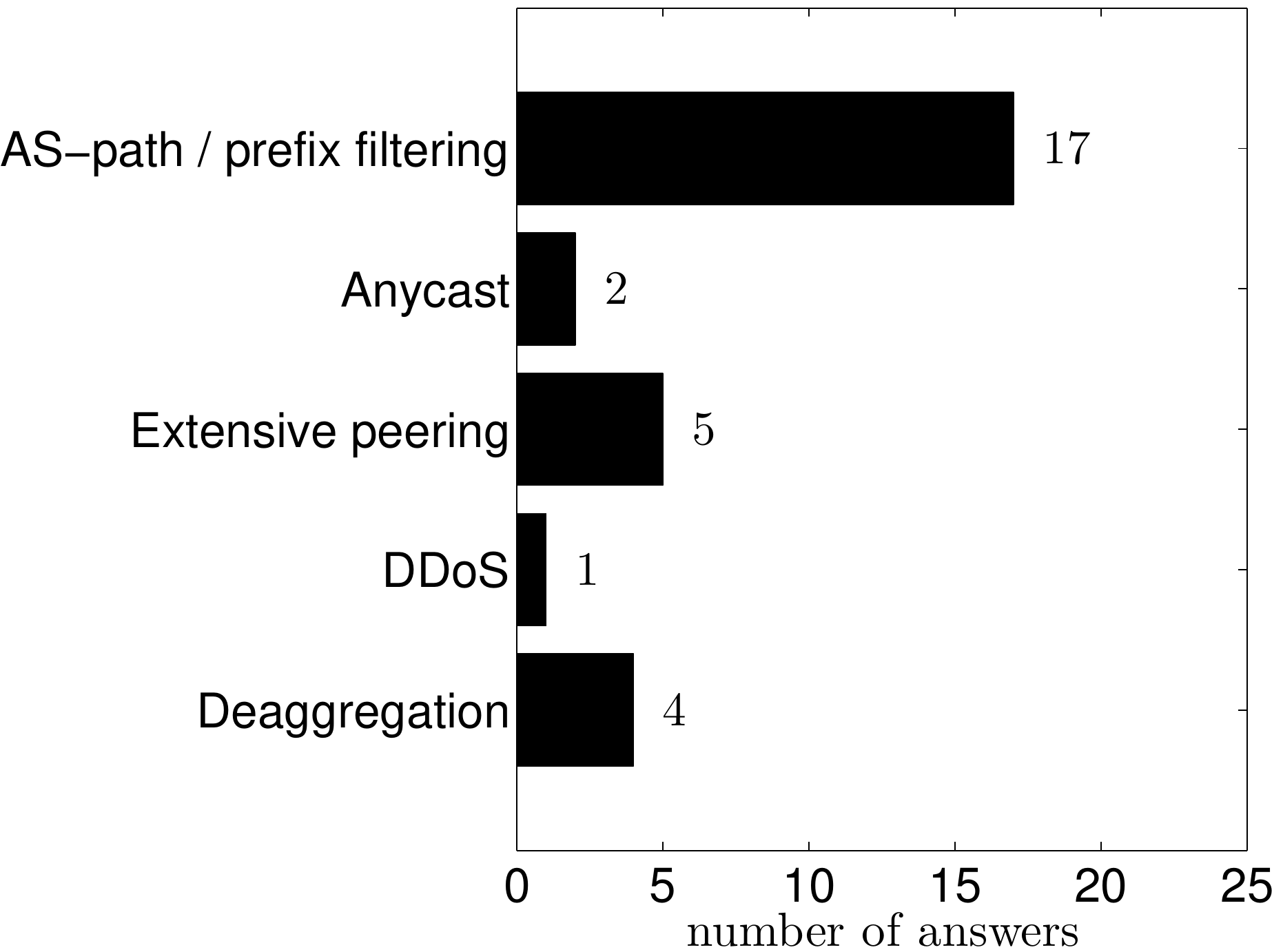}\label{fig:survey-q14}}
\end{minipage}

\vspace{0.3\baselineskip}

\begin{minipage}[b]{0.23\linewidth}
\subfigure[Q15: In your network, how would you learn about a hijacking incident against your prefix(es)?]{\includegraphics[width = 1\linewidth]{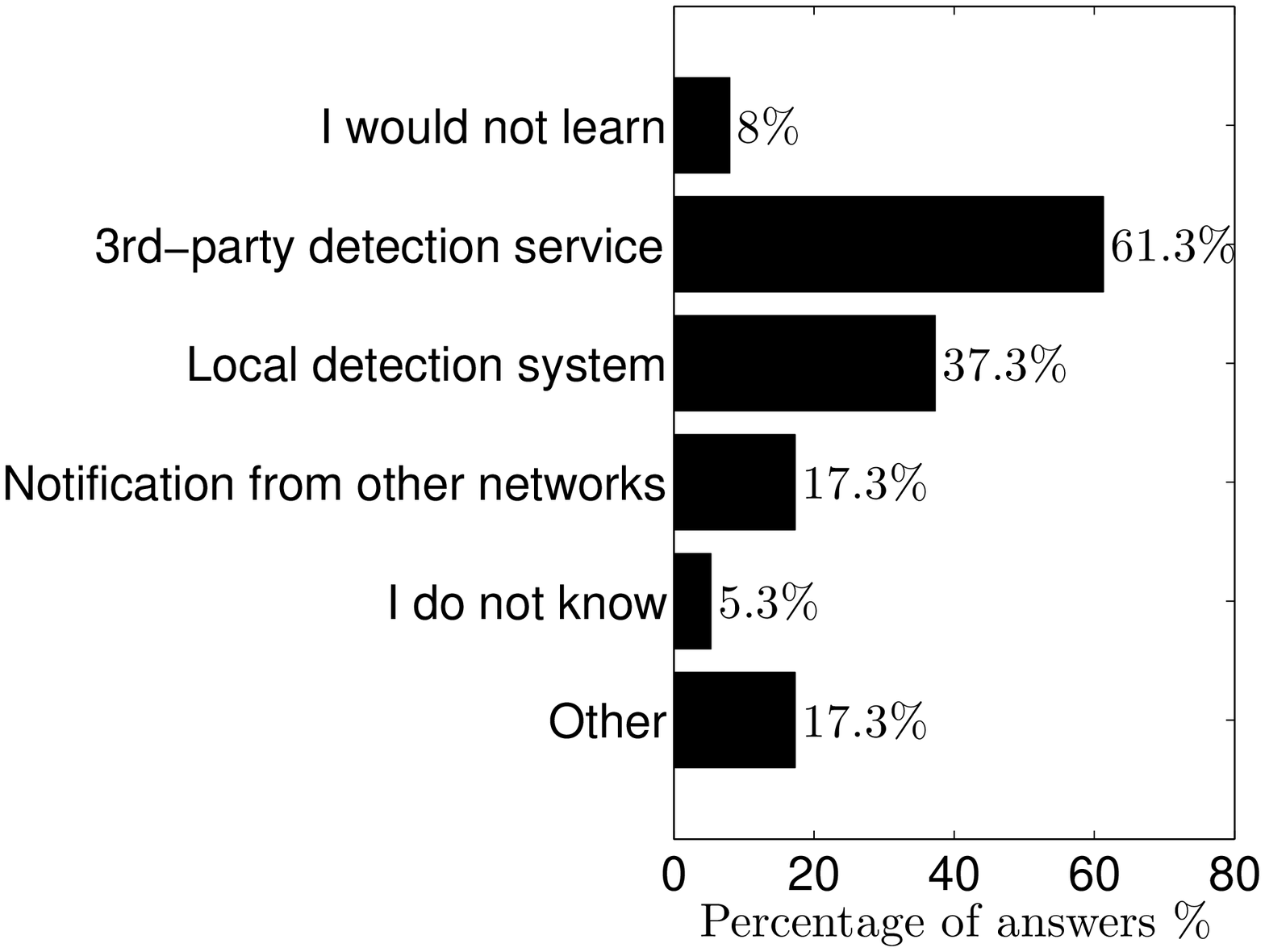}\label{fig:survey-q15}}
\end{minipage}
\hspace{0.1\linewidth}
\begin{minipage}[b]{0.23\linewidth}
\subfigure[Q16: If you use a local or third-party detection service or system, could you please give us more details about it? (open question/answers - optional)]{\includegraphics[width = 1\linewidth]{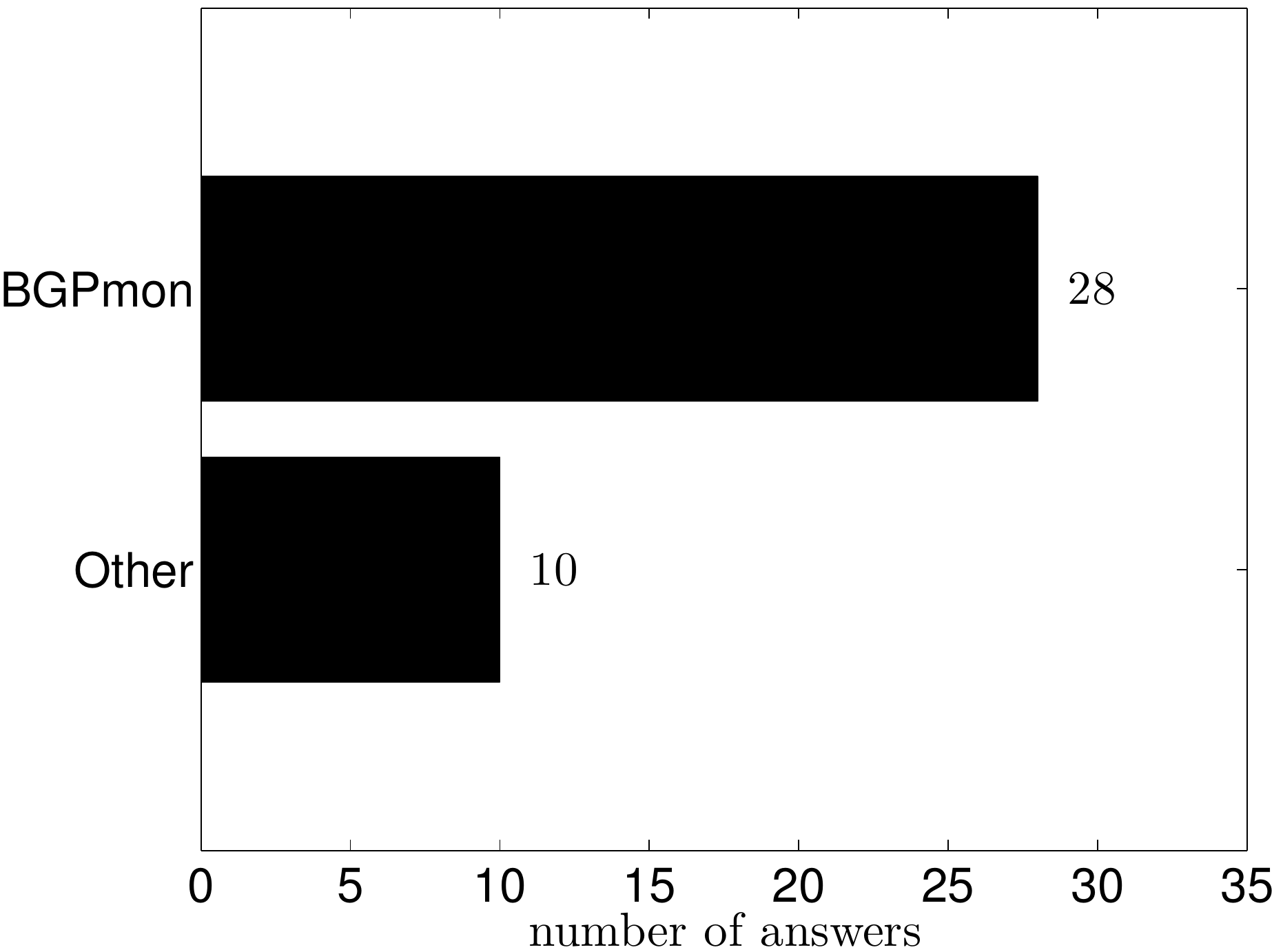}\label{fig:survey-q16}}
\end{minipage}
\hspace{0.1\linewidth}
\begin{minipage}[b]{0.23\linewidth}
\subfigure[Q17: How would you mitigate a hijack against your prefixes if you were notified about an on-going event?]{\includegraphics[width = 1\linewidth]{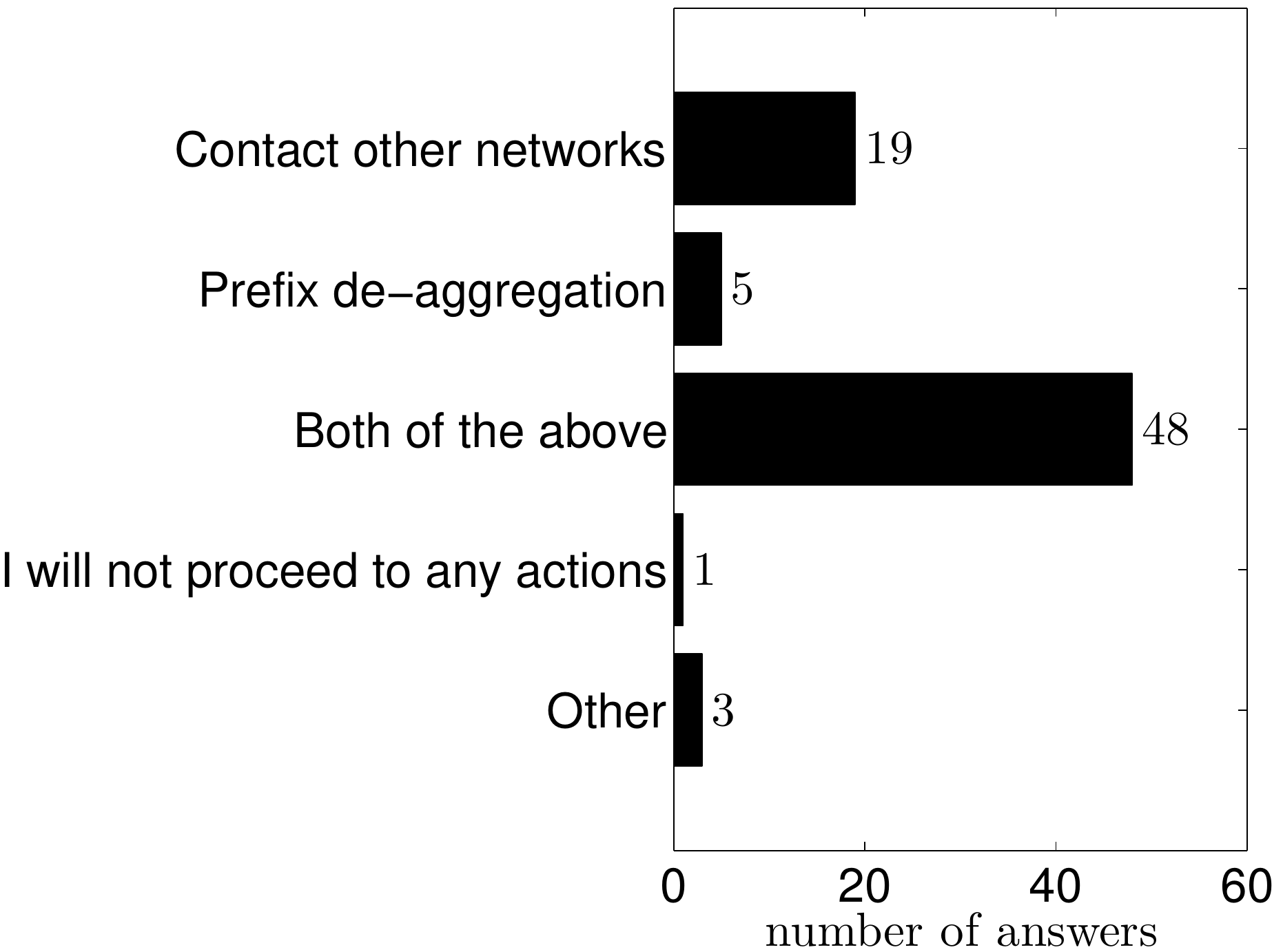}\label{fig:survey-q17}}
\end{minipage}

\vspace{0.3\baselineskip}

\begin{minipage}[b]{0.23\linewidth}
\subfigure[Q18: Would you outsource functions relating to the detection and mitigation of prefix hijacking incidents to a third-party, if this helps your organization reduce its risks?]{\includegraphics[width = 1\linewidth]{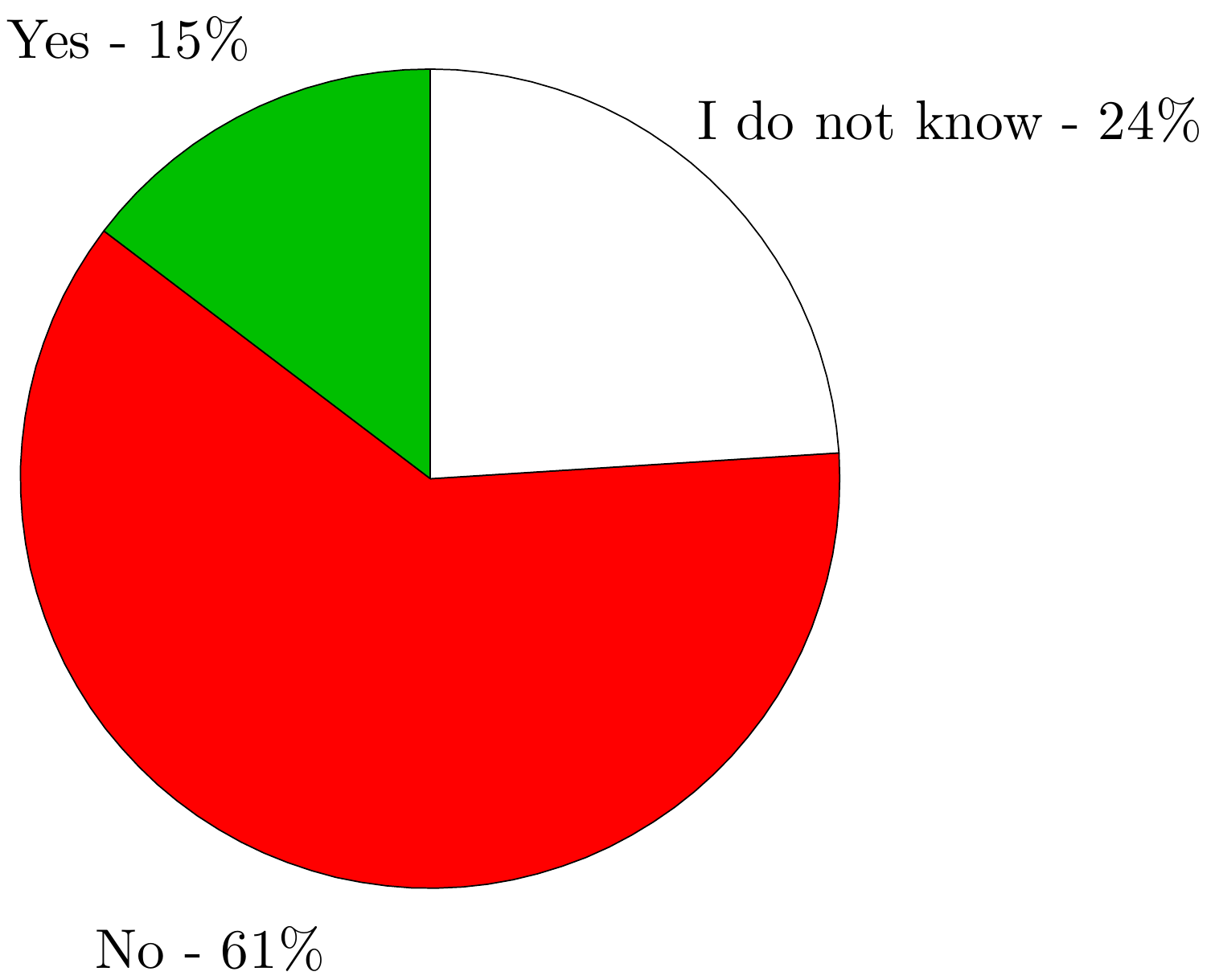}\label{fig:survey-q18}}
\end{minipage}
\hspace{0.01\linewidth}
\begin{minipage}[b]{0.23\linewidth}
\subfigure[Q19: If no (in Q18), what are the main factors that would affect your decision not to outsource prefix hijacking mitigation? (max 2 answers - optional)]{\includegraphics[width = 1\linewidth]{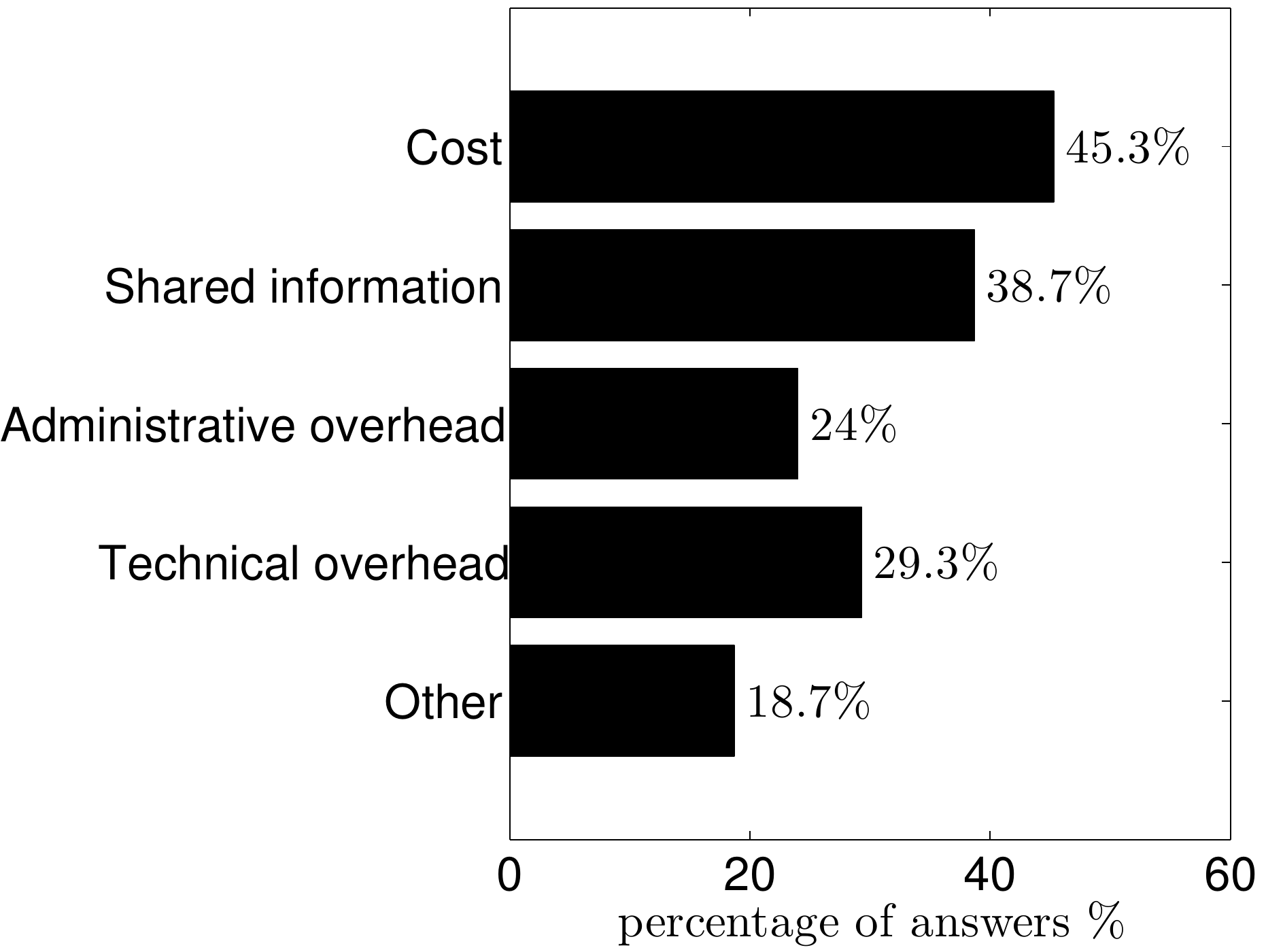}\label{fig:survey-q19}}
\end{minipage}
\hspace{0.01\linewidth}
\begin{minipage}[b]{0.23\linewidth}
\subfigure[Q20: Assuming you fully trust an outsourcing organization for prefix hijacking mitigation, what is the information/control (if any) you are still NOT willing to share/allow?]{\includegraphics[width = 1\linewidth]{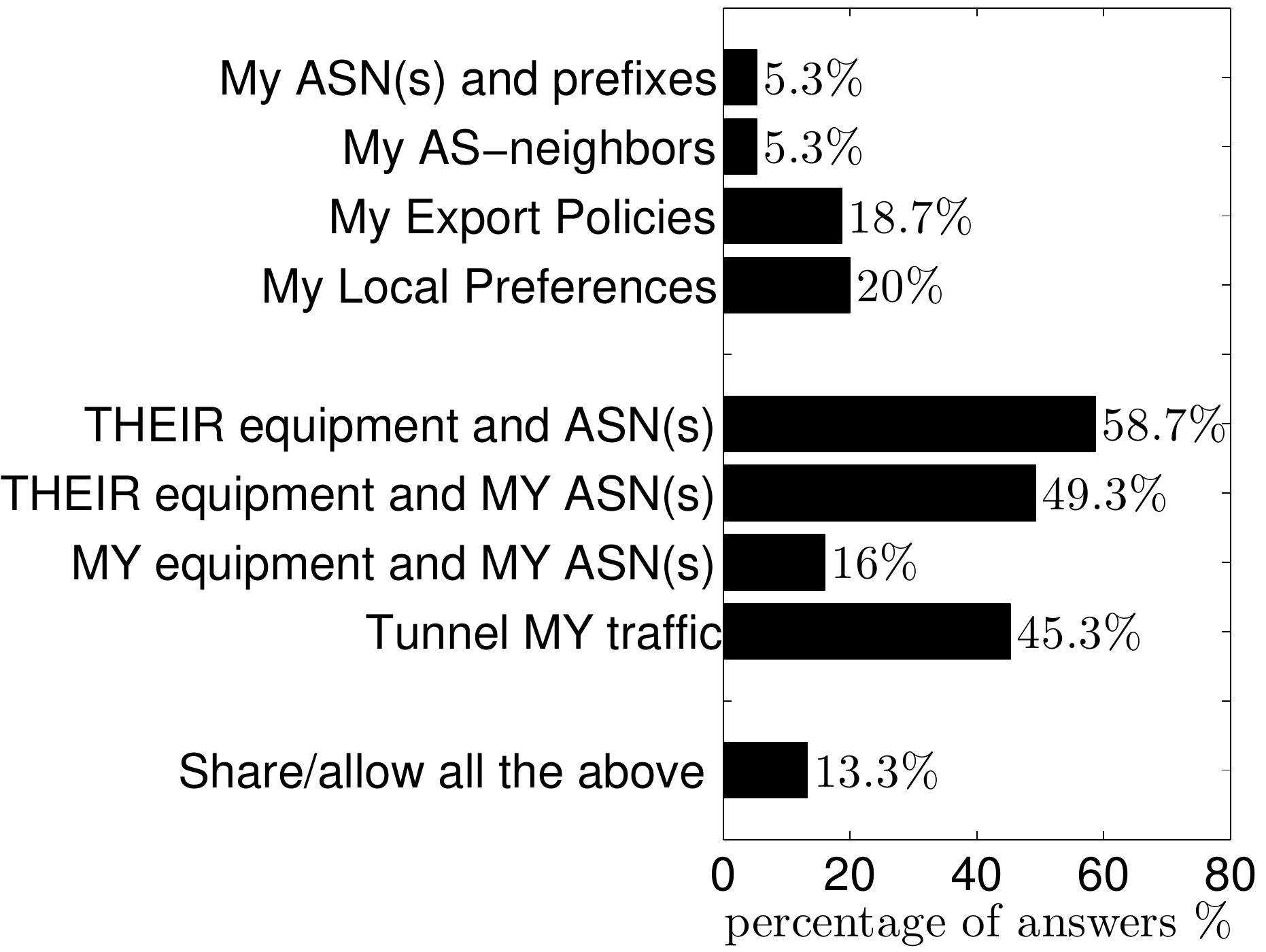}\label{fig:survey-q20}}
\end{minipage}
\hspace{0.01\linewidth}
\begin{minipage}[b]{0.23\linewidth}
\subfigure[Q21: How important do you consider the following characteristics for the deployment of a new defense system in your network?]{\includegraphics[width = 1\linewidth]{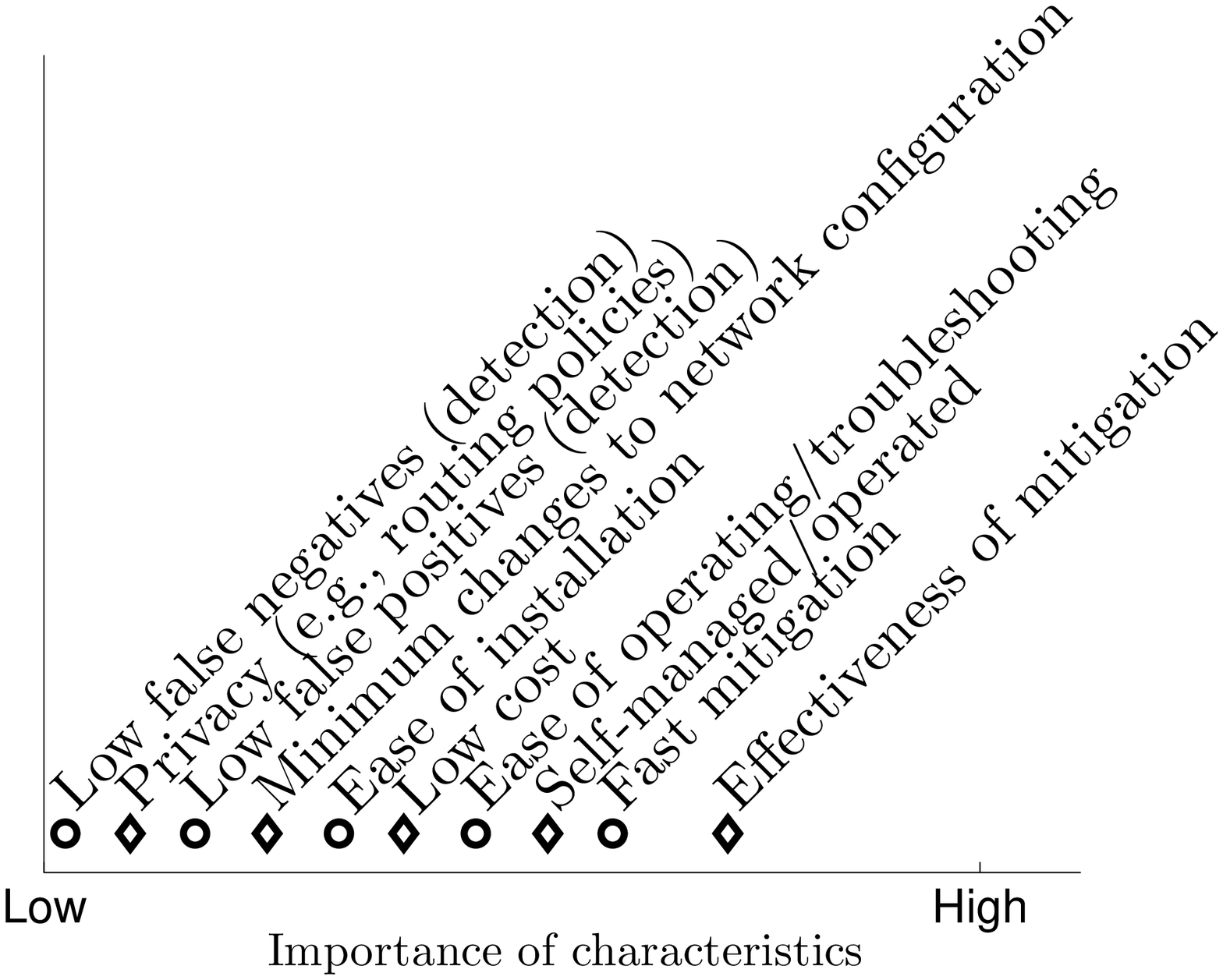}\label{fig:survey-q21}}
\end{minipage}


\begin{minipage}[b]{1\linewidth}
\centering
\subfigure[Detailed answers for Q21]{
\begin{small}
\begin{tabular}{|l|cccc|}
\hline
{\textbf{Importance (0: Low ... 3:High)}}& {\textbf{\cm{0}}}& {\textbf{\cm{1}}}& {\textbf{\cm{2}}}& {\textbf{\cm{3}}}\\
\hline
{Effectiveness of mitigation}				& {~0~~\%}& {~2.7\%}& {49.3\%}& {48.0\%}\\
\rowcolor{Gray}
{Fast mitigation}							& {~2.7\%}& {14.7\%}& {38.7\%}& {44.0\%}\\
{Self-managed/operated}						& {~1.3\%}& {18.7\%}& {38.7\%}& {41.3\%}\\
\rowcolor{Gray}
{Ease of operating/troubleshooting}			& {~0~~\%}& {21.3\%}& {40.0\%}& {38.7\%}\\
{Low cost}									& {~6.7\%}& {17.3\%}& {52.0\%}& {24.0\%}\\
\rowcolor{Gray}
{Ease of installation}						& {~1.3\%}& {30.7\%}& {53.3\%}& {14.7\%}\\
{Minimum changes to network configuration}	& {~5.3\%}& {26.7\%}& {49.3\%}& {18.7\%}\\
\rowcolor{Gray}
{Low false positives (detection)}			& {~2.7\%}& {12.0\%}& {29.3\%}& {41.3\%}\\
{Privacy (e.g., routing policies)}			& {~8.0\%}& {32.0\%}& {32.0\%}& {28.0\%}\\
\rowcolor{Gray}
{Low false negatives (detection)}			& {~1.3\%}& {13.3\%}& {36.0\%}& {34.7\%}\\
\hline
\end{tabular}
\end{small}
\label{table:survey-q21}
}
\end{minipage}

\caption{Survey results -- Defenses against BGP Prefix Hijacking}
\label{fig:questions-section-3}
\end{figure*}

The survey results show that the main practices that networks currently use for hijack mitigation comprise prefix de-aggregation and contacting other networks (Figs.~\ref{fig:survey-q14} and~\ref{fig:survey-q17}). Since these approaches have some important shortcomings, \eg de-aggregation is not efficient when a /24 prefix is hijacked (due to upstream filtering), and contacting network operators is usually done manually and thus adds significant delay to the mitigation process, we ask the network operators about their willingness to deploy new mitigation mechanisms, as well as what desired characteristics these mechanisms should possess.

We first ask them about their willingness to outsource functions related to the detection and mitigation of hijacks to a third party, in order to enhance their defenses. 61\% of the operators are not willing to proceed to such outsourcing practices (Fig.~\ref{fig:survey-q18}). This shows that a potential mechanism should not be entirely based on outsourcing, since this would not be acceptable by many networks. Flexible approaches that could be operated in two modes, \ie self-operated and outsourced, could be promising, since a significant percentage of 39\% does not reject the possibility to outsource such functions.

The reasons for the operators' reluctance to outsource are given in Fig.~\ref{fig:survey-q19}, where the associated (high) cost and the need to share private information about their network are the main factors. Administrative and technical overhead may also prevent outsourcing. This is a first indication about the characteristics of a potential defense system: low cost, privacy-preserving, and easy to operate and manage.

More specific results about what would be the information/control that they would \textit{not} be willing to share/allow with an outsourcing organization, are given in Fig.~\ref{fig:survey-q20}. As it can be seen, most of them are willing to share information about their prefixes and AS-neighbors (95\%), as well as their routing policies (80\%). A smaller percentage would allow BGP announcements to be controlled or implemented by the outsourcing organization.

Finally, according to operators, the importance of different characteristics that a hijack defense system should have, is shown in Fig.~\ref{table:survey-q21} (ranked from the highest to the lowest importance). A graphical representation of the importance of these characteristics for the network operators is given in Fig.~\ref{fig:survey-q21}, where the rightmost characteristics are considered of the highest importance. The speed and effectiveness of the mitigation stage, as well as the self-operability and low cost and management overhead, are the highest-ranked characteristics. Moreover, the detection stage is required to generate few false positives, which indicates the need for high levels of detection accuracy.

\section{Conclusion}\label{sec:conclusion}
In this work, to increase community understanding of existing BGP hijacking defenses
and the needs of network operators, we presented the results of a survey of 75 network
operators around the world. 

Through the survey, we verified our intuition that BGP prefix hijacking is a 
real threat and concerns the vast majority of network operators; in fact, hijacks can have a severe 
and lasting impact on their own networks. In the context of combatting such hijacks,
operators can use proactive or reactive techniques. On the one hand, proactive mechanisms, such as RPKI,
have gained extremely little traction for multiple reasons, including limited adoption and high
cost and complexity of deployment. On the other hand, practical reactive defenses such as contacting other networks, 
route filtering, extensive peering and prefix de-aggregation are usually preferred methods to mitigate hijacks; however,
each has its own significant limitations, ranging from very slow mitigation speeds (\eg contacting other operators) to
inefficient mitigation (\eg de-aggregation for /24 prefixes).

In terms of detection, we observe that operators mainly rely on third parties, such as BGPmon.
However, the level of satisfaction varies wildly across operators. Moreover, most of them are reluctant to perform
similar outsourcing for the mitigation of the hijacks themselves; in fact, there are mixed feelings about
the kind and amount of information they would be willing to disclose to the third party, as well as the involved
costs and technical and administrative overhead.
The speed and effectiveness of the mitigation stage, as well as the self-operability and low cost and management overhead, are of paramount importance; moreover, the detection stage is required to generate few false positives, mandating high levels of detection accuracy.
The findings of this survey could inform the design and implementation of new concepts and methodologies,
such as ARTEMIS~\cite{artemis-techrep-2017,ARTEMIS-Demo-Sigcomm-2016}, as well as more secure inter-domain routing protocols in general.

\section*{Acknowledgements}
This work was supported by the European Research Council grant agreement no. 338402, the National Science Foundation grant CNS-1423659, and the Department of Homeland Security (DHS) Science and Technology Directorate, Cyber Security Division (DHS S\&T/CSD) via contract number HHSP233201600012C.

%
\balance

\begin{comment}
\printbibliography
\end{comment}

\bibliographystyle{abbrv}
%
\balance
\bibliography{outages,rfc,artemis}

\begin{thebibliography}{10}

\bibitem{hijack-YouTube}
{\small{\url{https://www.ripe.net/publications/news/industry-developments/youtube-hijacking-a-ripe-ncc-ris-case-study}}}.

\bibitem{hijack-ChinaTelecom}
{\small{\url{http://www.bgpmon.net/chinese-isp-hijacked-10-of-the-internet/}}}.

\bibitem{hijack-BitCoins}
\url{https://www.wired.com/2014/08/isp-bitcoin-theft/}.

\bibitem{backconnect-hijack-2016}
{\small{\url{http://seclists.org/nanog/2016/Sep/122}}}.

\bibitem{iranian-hijack-2017}
{\small{\url{http://dyn.com/blog/iran-leaks-censorship-via-bgp-hijacks/}}}.

\bibitem{russian-hijack-2017}
{\small{\url{https://arstechnica.com/security/2017/04/russian-controlled-telecom-hijacks-financial-services-internet-traffic/
  }}}.

\bibitem{commercial-bgpmon}
{BGP}mon (commercial).
\newblock \url{http://www.bgpmon.net}.

\bibitem{google-survey-hijacking}
{Survey on BGP prefix hijacking}.
\newblock \url{http://tinyurl.com/hijack-survey}.

\bibitem{ripe-pakistan}
{YouTube Hijacking: A RIPE NCC RIS case study}.
\newblock
  \url{http://www.ripe.net/internet-coordination/news/industry-developments/youtube-hijacking-a-ripe-ncc-ris-case-study},
  March 2008.

\bibitem{ARTEMIS-Demo-Sigcomm-2016}
G.~Chaviaras, P.~Gigis, P.~Sermpezis, and X.~Dimitropoulos.
\newblock {ARTEMIS}: Real-time detection and automatic mitigation for bgp
  prefix hijacking.
\newblock In {\em Proc. ACM SIGCOMM Demo}, 2016.

\bibitem{cooper-2013-hotnets}
D.~Cooper, E.~Heilman, K.~Brogle, L.~Reyzin, and S.~Goldberg.
\newblock {On the Risk of Misbehaving RPKI Authorities}.
\newblock In {\em Proc. of ACM Workshop on Hot Topics in Networks
  (HotNets-XII)}, 2013.

\bibitem{george-2014-adventures}
W.~George.
\newblock {Adventures in RPKI (non) Deployment}.
\newblock
  \url{https://www.nanog.org/sites/default/files/wednesday_george_adventuresinrpki_62.9.pdf},
  2014.
\newblock NANOG presentation.

\bibitem{RPKI-deployment-2016}
Y.~Gilad, A.~Cohen, A.~Herzberg, M.~Schapira, and H.~Shulman.
\newblock Are we there yet? on {RPKI}'s deployment and security.
\newblock NDSS 2017, to appear, \url{http://eprint.iacr.org/2016/1010.pdf},
  2016.

\bibitem{Goldberg-why-so-long-ACM-Comm-2014}
S.~Goldberg.
\newblock Why is it taking so long to secure internet routing?
\newblock {\em Communications of the ACM}, 57(10):56--63, 2014.

\bibitem{BGPv4}
S.~Hares, Y.~Rekhter, and T.~Li.
\newblock A border gateway protocol 4 (bgp-4).
\newblock \url{https://tools.ietf.org/html/rfc4271}, 2006.

\bibitem{Karlin-PGBGP-ICNP-2006}
J.~Karlin, S.~Forrest, and J.~Rexford.
\newblock Pretty good bgp: Improving bgp by cautiously adopting routes.
\newblock In {\em Proc. IEEE ICNP}, 2006.

\bibitem{Kent-secure-BGP-JSAC-2000}
S.~Kent, C.~Lynn, and K.~Seo.
\newblock Secure border gateway protocol (s-bgp).
\newblock {\em IEEE Journal on Selected Areas in Communications},
  18(4):582--592, 2000.

\bibitem{bgpsec-specification-2015}
M.~Lepinski.
\newblock Bgpsec protocol specification.
\newblock 2015.

\bibitem{rpki-rfc}
M.~Lepinski, R.~Barnes, and S.~Kent.
\newblock An infrastructure to support secure internet routing.
\newblock 2012.

\bibitem{lychev-2013-sigcomm}
R.~Lychev, S.~Goldberg, and M.~Schapira.
\newblock {BGP Security in Partial Deployment: Is the Juice Worth the Squeeze?}
\newblock In {\em Proc. of ACM SIGCOMM}, 2013.

\bibitem{matsumoto-2017-privsec}
S.~Matsumoto, R.~M. Reischuk, P.~Szalachowski, T.~H.-J. Kim, and A.~Perrig.
\newblock {Authentication Challenges in a Global Environment}.
\newblock {\em ACM Trans. Priv. Secur.}, 20:1:1--1:34, 2017.

\bibitem{Ramachandran-BGP-spammers-CCR-2006}
A.~Ramachandran and N.~Feamster.
\newblock Understanding the network-level behavior of spammers.
\newblock {\em ACM SIGCOMM Computer Communication Review}, 36(4):291--302,
  2006.

\bibitem{artemis-techrep-2017}
P.~Sermpezis, V.~Kotronis, P.~Gigis, X.~Dimitropoulos, J.~H. Park, D.~Cicalese,
  A.~King, and A.~Dainotti.
\newblock {ARTEMIS: Neutralizing BGP Hijacking within a Minute}.
\newblock \url{arxiv.org/abs/1801.01085}, 2017.
\newblock {arXiv 1801.01085}.

\bibitem{Subramanian-listen-whisper-NSDI-2004}
L.~Subramanian, V.~Roth, I.~Stoica, S.~Shenker, and R.~Katz.
\newblock Listen and whisper: Security mechanisms for bgp.
\newblock In {\em Proc. NSDI}, 2004.

\bibitem{Vervier-mind-blocks-NDSS-2015}
P.-A. Vervier, O.~Thonnard, and M.~Dacier.
\newblock Mind your blocks: On the stealthiness of malicious bgp hijacks.
\newblock In {\em Proc. NDSS}, 2015.

\end{thebibliography}

%
%
%
%

%
%
%
%

\end{document}